\documentclass[12pt,notitlepage,aps,prl,amsmath,amssymb]{revtex4-1}

\usepackage{bm}
\usepackage{graphicx}

\usepackage{times}
\usepackage{mathptmx}

\newcommand{\wh}{\widehat}

\newcommand{\ave}[1]{{\left<#1\right>}}

\newcommand{\ibid}{\textit{ibid}.}
\newcommand{\etal}{\textit{et al.}}

\newcommand{\sol}{{SOL}}

\newcommand{\Phirms}{\ensuremath{\Phi_\text{rms}}}
\newcommand{\taud}{\ensuremath{\tau_\text{d}}}
\newcommand{\tauw}{\ensuremath{\tau_\text{w}}}

\newcommand{\Eqref}[1]{Eq.~\eqref{#1}}

\newcommand{\Figref}[1]{Fig.~\ref{#1}}

\newcommand{\CJP}{\textit{Czech.\ J.\ Phys.}}
\newcommand{\JNM}{\textit{J.~Nuclear Mater.}}

\newcommand{\NF}{\textit{Nucl.\ Fusion}}

\newcommand{\PPCF}{\textit{Plasma Phys.\ Contr.\ Fusion}}
\newcommand{\PFR}{\textit{Plasma Fusion Res.}}

\newcommand{\PP}{\textit{Phys.\ Plasmas}}
\newcommand{\PS}{\textit{Phys.\ Scripta}}
\newcommand{\PRE}{\textit{Phys.\ Rev.~E}}

\newcommand{\BTSJ}{\textit{Bell Sys.\ Tech. J.}}
\newcommand{\PCPS}{\textit{Proc.\ Cambridge Phil.\ Soc.}}
\newcommand{\JPO}{\textit{J.~Phys.\ Ocean.}}
\newcommand{\PRL}{\textit{Phys.~Rev.\ Lett.}}

\begin{document}

\title{Stochastic modelling of intermittent scrape-off layer plasma fluctuations}

\author{O.~E.~Garcia}

\email{odd.erik.garcia@uit.no}

\affiliation{Department of Physics and Technology, University of Troms{\o}, N-9037 Troms{\o}, Norway}
\affiliation{MIT Plasma Science and Fusion Center, Cambridge, Massachusetts 02139, USA}

\date{May 24, 2012}

\begin{abstract}
Single-point measurements of fluctuations in the scrape-off layer
of magnetized plasmas are generally found to be dominated by
large-amplitude bursts which are associated with radial motion
of blob-like structures. A stochastic model for these fluctuations
is presented, with the plasma density given by a random sequence
of bursts with a fixed wave form. Under very general conditions,
this model predicts a parabolic relation between the skewness and
kurtosis moments of the plasma fluctuations. In the case of
exponentially distributed burst amplitudes and waiting times,
the probability density function for the fluctuation amplitudes is
shown to be a Gamma distribution with the scale parameter given
by the average burst amplitude and the shape parameter given by
the ratio of the burst  duration and waiting times.
\end{abstract}

\maketitle

Cross-field transport of particles and heat in the scrape-off layer (\sol)
of non-uniformly magnetized plasmas is caused by radial motion of
blob-like structures \cite{zweben,garcia-pfr,dmz,terry,kube,garcia-blob}.
This results in single-point recordings dominated by large-amplitude
bursts, which have an asymmetric wave form with a fast rise and a slow
decay, and positively skewed and flattened amplitude probability density
functions \cite{boedo,garcia-tcv-ppcf3,graves,garcia-tcv-ppcf1,garcia-tcv-ppcf2,garcia-tcv-nf,garcia-esel,labombard1}.
Measurements on a number of tokamak experiments have
demonstrated that as the empirical discharge density is approached,
the radial \sol\ particle density profile becomes broader and plasma--wall
interactions increase
\cite{graves,labombard1,lipschultz,garcia-tcv-nf,garcia-tcv-ppcf2,labombard1}.

Probe measurements on Tokamak {\`a} Configuration Variable (TCV) have
demonstrated a remarkable degree of universality of the plasma fluctuations
in the far \sol\ region, which is dominated by radial motion of filament
structures and a relative fluctuation level of order unity
\cite{graves,garcia-tcv-ppcf1,garcia-tcv-nf,garcia-tcv-ppcf2}.
In particular, the amplitude
distribution of the plasma fluctuations are found to
be well described by a Gamma distribution across a broad range of
plasma parameters and for all radial positions in the \sol\ \cite{graves}. Excellent agreement was found when comparing analysis of these data
with turbulence simulations based on interchange motions
\cite{garcia-tcv-nf,garcia-tcv-ppcf1,garcia-esel}.

In this Letter,
a stochastic model for intermittent fluctuations in the plasma \sol\
is presented with all statistical properties in agreement with experimental
measurements. It is demonstrated that this model explains many of the
salient experimental findings and empirical scaling relations, including
broad plasma profiles and large fluctuation levels, skewed and flattened
amplitude probability distribution functions and a parabolic relation
between the skewness and kurtosis moments. The latter has been
observed in the boundary region of numerous experiments on
magnetized plasmas as well as in hydrodynamical and astrophysical systems dominated by intermittent fluctuations \cite{graves,labit,sattin,garcia-psi}.

There have been several previous attempts at describing the universal
features of intermittent fluctuations at the boundary of magnetically
confined plasmas \cite{sandberg,sattin2,krommes}. However, none
of these models provide the appealing simplicity, physical insight,
novel predictions and favourable comparison with experimental
measurements as the theory presented here. In particular, the
statistical properties which the present model is based on have
been directly confirmed by experiments
\cite{garcia-tcv-ppcf3,garcia-psi}.

Experimental measurements as well as numerical simulations suggest
that plasma fluctuations in the far \sol\ can be represented as a random
sequence of bursts events,
\begin{equation} \label{shotnoise}
\Phi(t) = \sum_k A_k\psi(t-t_k)
\end{equation}
where $A_k$ is the amplitude and $t_k$ is the arrival time
for burst event $k$, and $\psi$ is a fixed burst wave form.
This stochastic process resembles a general class of
models known as ``shot noise'', in which the noise is generated by the
addition of a large number of disturbances \cite{campbell,rice,pecseli}.
The objective is to estimate the mean value and higher order moments
of $\Phi$, the amplitude probability density function $P_\Phi$ and
discuss how the burst statistics are related to broad \sol\ plasma
profiles and large fluctuation levels.

If there are $K$ burst events in a time interval $T$, the average burst
waiting time $\tauw$ is given by $T/K$. It follows that the mean
value of the plasma density is \cite{campbell,rice,pecseli},
\begin{equation} \label{phiave}
\ave{\Phi} = \frac{\ave{A}}{\tauw}\,\int_{-\infty}^{\infty} dt\,\psi(t) .
\end{equation}
Here and in the following, angular brackets are defined as an average
of a random variable over all its values. The above equation shows that
the mean plasma density is given by the average burst amplitude and
the ratio of the burst duration and waiting times. Equation~\eqref{phiave}
thus elucidates the role of burst statistics for high plasma density in the
far \sol. It should be noted that this result only depends on the integrated
burst wave form and the average burst amplitude and waiting time.

Expressions for the variance and higher order moments of $\Phi$
have been derived in the case that burst events occur in accordance
to a Poisson process with rate $1/\tauw$. The probability of exactly
$K$ burst events in a time interval $T$ is then given by the Poisson
distribution,
\begin{equation} \label{poisson}
P(K) = \exp\left(-\frac{T}{\tauw} \right)\left(\frac{T}{\tauw}\right)^K
\frac{1}{K!} .
\end{equation}
From this it follows that the burst waiting times are exponentially
distributed, as found from experimental measurements
\cite{garcia-tcv-ppcf3,garcia-psi},
\begin{equation} \label{expwait}
P_\tau(\tau) = \frac{1}{\tauw}\,\exp\left( -\frac{\tau}{\tauw} \right) .
\end{equation}
The general result states that for the stochastic process defined by
 \Eqref{shotnoise}, the cumulants $\kappa_n$ for the probability
density $P_\Phi(\Phi)$ are given by \cite{rice,pecseli}
\begin{equation} \label{cumulant}
\kappa_n = \frac{\ave{A^n}I_n}{\tauw} ,
\end{equation}
where the integral of the $n$-th power of the wave form is defined by
\begin{equation} \label{taud}
I_n = \int_{-\infty}^{\infty} dt\,\left[\psi(t)\right]^n .
\end{equation}
The cumulants are the coefficients in the expansion of the logarithm of
the characteristic function for $P_\Phi$,
\begin{equation} \label{lncfe}
\ln \ave{\exp(i\Phi u)} =
\sum_{n=1}^{\infty} \kappa_n\,\frac{(iu)^n}{n!} .
\end{equation}
A power series expansion shows that the characteristic function
is related to the raw moments of $\Phi$, defined by $\mu_n'=\ave{\Phi^n}$,
\begin{equation} \label{cfs}
\ave{\exp(i\Phi u)} = 1 + \sum_{n=1}^{\infty} \frac{\ave{i\Phi u}^n}{n!} =
1 + \sum_{n=1}^{\infty} \mu_n'\,\frac{(iu)^n}{n!} .
\end{equation}
Further expanding the logarithmic function in \Eqref{lncfe} and
using \Eqref{cfs},  it follows that the lowest order centred moments
$\mu_n=\ave{(\Phi-\ave{\Phi})^n}$ are related to the cumulants
by the relations $\mu_2=\kappa_2$, $\mu_3=\kappa_3$ and
$\mu_4=\kappa_4+3\kappa_2^2$.

The variance and higher order moments are straight forward to
calculate from \Eqref{cumulant} for general burst wave forms and
amplitude distributions. The coefficient of variation, skewness and
flatness are given respectively by
\begin{subequations} \label{moments}
\begin{align}
C & = \frac{\ave{(\Phi-\ave{\Phi})^2}^{1/2}}{\ave{\Phi}}  =
\tauw^{1/2}\,\frac{I_2^{1/2}}{I_1}\,\frac{\ave{A^2}^{1/2}}{\ave{A}} , \\
S & = \frac{\ave{(\Phi-\ave{\Phi})^3}}{\Phirms^3} =
\tauw^{1/2}\,\frac{I_3}{I_2^{3/2}}\frac{\ave{A^3}}{\ave{A^2}^{3/2}} , \\
F & = \frac{\ave{(\Phi-\ave{\Phi})^4}}{\Phirms^4} =
3 + \tauw\,\frac{I_4}{I_2^2}\frac{\ave{A^4}}{\ave{A^2}^2} .
\end{align}
\end{subequations}
The two latter relations imply that there is a parabolic relation
between the skewness and flatness moments,
\begin{equation} \label{parabolic}
F = 3 + \frac{I_2I_4}{I_3^2}\frac{\ave{A^2}\ave{A^4}}{\ave{A^3}^2}\,S^2 .
\end{equation}
Such a parabolic relation between the third and fourth order moments
have been found for a wide variety of physical systems dominated by
intermittent fluctuations
\cite{graves,garcia-psi,labit,sattin,sandberg,sattin2,krommes}.

The expressions for the higher order moments become particularly simple
for a burst wave form given by a sharp rise followed by a slow exponential decay,
\begin{equation} \label{expburst}
\psi(t) = \Theta(t) \exp\left( - \frac{t}{\taud} \right) , 
\end{equation}
where $\Theta$ is the step function and $\taud$ is the burst
duration time. This is the typical wave form found from probe
and gas puff imaging measurements in the far \sol\
\cite{boedo,garcia-tcv-ppcf3,graves,garcia-tcv-ppcf1,garcia-tcv-ppcf2,garcia-tcv-nf,garcia-esel}.
The integral given in \Eqref{taud} is then $I_n=\taud/n$ and the
cumulants are thus given by $\kappa_n=\taud\ave{A^n}/n\tauw$.
The expressions for the coefficient of variation, skewness and
flatness become
\begin{subequations}
\begin{align} \label{momentsexpw}
C & = \left( \frac{\tauw}{2\taud} \right)^{1/2}\frac{\ave{A^2}^{1/2}}{\ave{A}} , \\
S & = \left( \frac{8\tauw}{9\taud} \right)^{1/2}\frac{\ave{A^3}}{\ave{A^2}^{3/2}} , \\
F & = 3 + \frac{\tauw}{\taud}\frac{\ave{A^4}}{\ave{A^2}^2} .
\end{align}
\end{subequations}
The relation between the skewness and flatness is in this case given by
\begin{equation}
F = 3 + \frac{9}{8}\frac{\ave{A^2}\ave{A^4}}{\ave{A^3}^2}\,S^2 .
\end{equation}
Note that independent of the burst wave form, the probability distribution
function for $\Phi$ is positively skewed, $S>0$, and flattened, $F>3$, for
positive definite burst amplitudes $A$.

The above expressions for the lowest order moments simplify
further in the case of exponentially distributed burst amplitudes,
\begin{equation} \label{expamp}
P_A(A) = \frac{1}{\ave{A}}\,\exp\left( -\frac{A}{\ave{A}} \right) .
\end{equation}
which is also  consistent with experimental measurements in the
\sol\ of magnetically confined plasmas \cite{garcia-tcv-ppcf3,garcia-psi}.
The raw amplitude moments are then given by
$\ave{A^n} = \ave{A}^nn!$. In this case the relative fluctuation level,
skewness and flatness can be written as
\begin{equation}
C = \left( \frac{\tauw}{\taud} \right)^{1/2} \qquad
S = \left( \frac{4\tauw}{\taud} \right)^{1/2} , \qquad
F = 3 + \frac{6\tauw}{\taud} .
\end{equation}
All these moments increase with the ratio $\tauw/\taud$. The parameter
$\gamma=\taud/\tauw$ is thus a measure of intermittency in
the shot noise process. The relation between the skewness and
flatness moments now becomes
\begin{equation}
F = 3 + \frac{3}{2}\,S^2 ,
\end{equation}
which is in excellent agreement with measurements in the
\sol\ of tokamak plasmas \cite{garcia-psi,graves,sattin}.

\begin{figure}
\centering
\includegraphics[width=10cm]{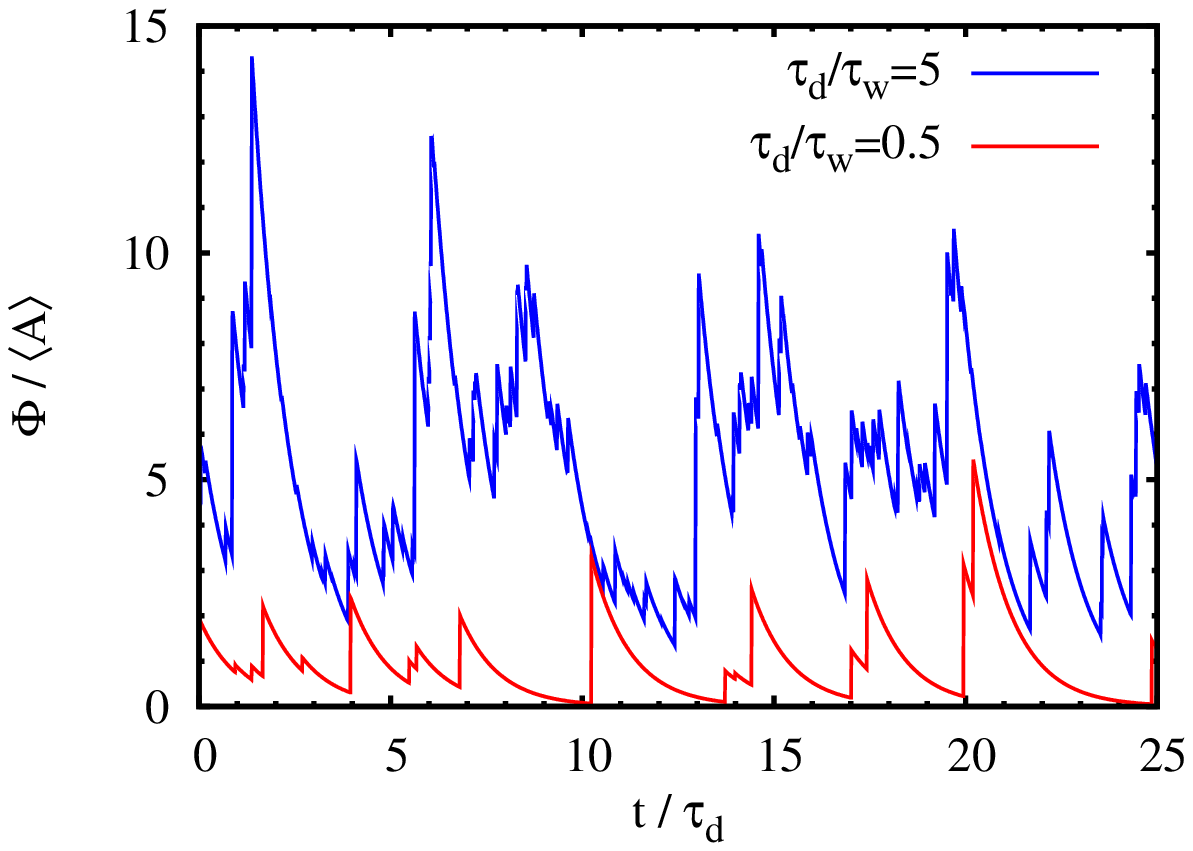}
\caption{Simulated shot noise time series for exponentially
distributed burst amplitudes and waiting times.}
\label{sgnl}
\end{figure}

The interpretation of these results is evident. For short waiting times and
long burst duration, the signal $\Phi$ will at any time be influenced by
many individual bursts, resulting in a large mean value and small relative
variation. In the opposite limit of long waiting times and short duration, the
signal is dominated by isolated burst events, resulting in a smaller mean value
and large relative fluctuations, skewness and flatness. This is clearly
illustrated in \Figref{sgnl}, which shows two numerical examples of the
shot noise process given by \Eqref{shotnoise} for exponentially distributed
burst waiting times and amplitudes.

The results presented above show that the skewness and flatness
vanish in the limit of large $\gamma$. It can be demonstrated that
the probability density function for $\Phi$ then approaches a
normal distribution \cite{rice,pecseli}. The distribution $P_\Phi$
can be written in terms of the characteristic function given in \Eqref{lncfe},
\begin{equation}
P_\Phi(\Phi) = \frac{1}{2\pi}\int_{-\infty}^{\infty} du\,\exp\left[
- i\Phi u + \sum_{n=1}^\infty \frac{(iu)^n\kappa_n}{n!} \right] .
\end{equation}
In the limit of small $\tauw/\taud$ the exponential function can be
expanded as a power series in $u$. Integrating term by term then gives
\begin{equation} \label{Gausslimit}
\Phirms\,P_\Phi(\Phi) =
\frac{1}{(2\pi)^{1/2}}\,\exp\left( -\frac{\wh{\Phi}^2}{2} \right) \left[ 1 +
 \frac{\mu_3}{3!\Phirms^3(2\pi)^{1/2}}( \wh{\Phi}^3 - 3\wh{\Phi} ) +
R(\wh{\Phi}) \right] ,
\end{equation}
where $R$ is the sum of the remaining terms in the expansion
and the centred and rescaled amplitude is defined by
$\wh{\Phi}=(\Phi-\ave{\Phi})/\Phirms$. The terms
inside the square bracket in \Eqref{Gausslimit} are of order 1,
$(\tauw/\taud)^{1/2}$ and $\tauw/\taud$, respectively. This result
shows how the probability density function for $\Phi$ approaches the
normal distribution in the limit of large $\gamma$. This transition
to normal distributed fluctuations is expected from the central limit
theorem, since in this case a large number of burst events contributes
to $\Phi$ at any given time.

The asymptotic probability density function in the strong intermittency
regime for small $\gamma$ can be obtained by neglecting overlap of
individual burst events. Considering first a single burst event
$\phi(t)=A\exp(-t/\taud)$, the time $dt$ spent between $\phi$ and
$\phi+d\phi$ is given by $dt/d\phi=\taud/\phi$.
Note that due to the assumed exponential wave form, the burst amplitude
$A$ does not enter this expression. The number of bursts with amplitude
above $\Phi$ is given by the complimentary cumulative amplitude
distribution function, which for an exponential distribution is 
$\exp(-\Phi/\ave{A})$. The probability density
function $P_\Phi$ is given by the proportion of time which $\Phi(t)$
spends in the range from $\Phi$ to $\Phi+d\Phi$. With the appropriate
normalization, the asymptotic probability density function in the strong
intermittency regime is thus given by
\begin{equation}
\lim_{\gamma\rightarrow0} P_\Phi(\Phi) = \lim_{\gamma\rightarrow0}
\frac{1}{\Gamma(\gamma)}\frac{1}{\Phi}\,\exp
\left( -\frac{\Phi}{\ave{A}} \right) ,
\end{equation}
where we have defined the Gamma-function
\begin{equation}
\Gamma(\gamma) =
\int_{0}^{\infty} d\varphi\,\varphi^{\gamma-1}\exp(-\varphi) ,
\end{equation}
which to lowest order is given by $1/\gamma$ in the limit of small
$\gamma$. This probability density function has an exponential
tail for large amplitudes but is inversely proportional to $\Phi$ for
small amplitudes due to the long quite period between burst events
in this strong intermittency regime.

The characteristic function for a sum of independent random variables
is the product of their individual characteristic functions. Thus, the
probability that a sum of $K$ burst events $\phi_k$ lies in the
range between $\Phi$ and $\Phi+d\Phi$ is given by
\begin{equation} \label{prbK}
\frac{d\Phi}{2\pi}\int_{-\infty}^{\infty} du\,\exp(-i\Phi u)
\prod_{k=1}^K\ave{\exp(i\phi_k u)} ,
\end{equation}
where the characteristic functions are averaged over the values
of $\phi_k$. For general amplitude distribution and burst wave
forms,
\begin{equation*}
\ave{\exp(i\phi_k u)} = \frac{1}{T}\int_0^T
dt_k\,\int_{-\infty}^{\infty} dA\,P_A(A)\exp\left[ iAu\psi(t-t_k)\right] ,
\end{equation*}
where $T$ is the duration of the time interval under consideration.
Since all the $K$ characteristic functions in \Eqref{prbK} are the same,
the conditional probability $P_K$ is given by
\begin{equation*}
P_K(\Phi) = \frac{1}{2\pi}\int_{-\infty}^{\infty} du\,\exp(-i\Phi u)
\ave{\exp(i\phi_k u)}^K ,
\end{equation*}
assuming the number of events $K$ in a time interval $T$ to
be given. The probability density function for the amplitude $\Phi$ is
given by summing over all $K$,
\begin{equation}
P_\Phi(\Phi) = \sum_{K=0}^{\infty}P(K)P_K(\Phi) ,
\end{equation}
where $P(K)$ is given by \Eqref{poisson}
The stationary probability density function for $\Phi$ is obtained by
letting  $T\rightarrow\infty$. Some elementary manipulations lead to
the desired result,
\begin{equation} \label{Pphig}
P_\Phi(\Phi) = \frac{1}{2\pi}\int_{-\infty}^{\infty} \exp\left\{ -i\Phi u +
\frac{1}{\tauw}\int_{-\infty}^{\infty} dA\,P_A(A) \int_{-\infty}^{\infty}
dt\,\left[ \exp(iAu\psi(t))-1 \right] \right\} .
\end{equation}
The logarithm of the characteristic function of $P_\Phi$ is thus
\begin{equation} \label{lncf2}
\frac{1}{\tauw} \int_{-\infty}^{\infty} dA\,P_A(A) \int_{-\infty}^{\infty}
dt\,\left[ \exp(iAu\psi(t))-1 \right] = \sum_{n=1}^{\infty}
\frac{1}{\tauw}\frac{(iu)^n}{n!} \int_{-\infty}^{\infty}
dA\,A^nP_A(A) \int_{-\infty}^{\infty} dt\,[\psi(t)]^n .
\end{equation}
where again the exponential function has been expanded. This
establishes the general result stated by \Eqref{cumulant}.

For the special case of exponentially distributed burst amplitudes,
the amplitude integral in the above equation is given by $\ave{A}^nn!$,
cancelling the factorial in \Eqref{lncf2}. Further invoking the
exponential wave form given in \Eqref{expburst}, it follows that
the characteristic function for the stationary distribution can be
written as
\begin{equation}
\exp\left[ \gamma\sum_{n=1}^{\infty} \frac{(i\ave{A}u)^n}{n} \right] =
\left( 1-i\ave{A}u \right)^{-\gamma} .
\end{equation}
This is nothing but the characteristic function for a Gamma distribution
with scale parameter $\ave{A}$ and shape parameter $\gamma$. Thus,
the probability density function for $\Phi$ is given by
\begin{equation} \label{gdfa}
P_\Phi(\Phi) = \frac{1}{\ave{A}\Gamma(\gamma)}
\left( \frac{\Phi}{\ave{A}} \right)^{\gamma-1}
\exp\left( - \frac{\Phi}{\ave{A}} \right) .
\end{equation}
The lowest order moments and asymptotic limits of this distribution
agree with the expressions discussed previously.  In particular, the mean
value is given by $\ave{\Phi}=\gamma\ave{A}$ and the variance by 
$\Phirms^2=\gamma\ave{A}^2$.

For $\gamma>1$, the most likely amplitude of the Gamma distribution
is $(\gamma-1)\ave{A}$ and the shape of the distribution function is
unimodal and skewed. When $\gamma=1$, $P_\Phi$
becomes an exponential distribution with the mean density given
by the average burst amplitude,
\begin{equation} \label{phiexp}
P_\Phi(\Phi) = \frac{1}{\ave{\Phi}}\,\exp\left( - \frac{\Phi}{\ave{\Phi}} \right) .
\end{equation}
Note that by writing the average burst amplitude as
$\ave{A}=\ave{\Phi}/\gamma$,  the Gamma distribution given
in \Eqref{gdfa} can be written in terms of the average plasma density as
\begin{equation} \label{gamman}
\ave{\Phi}P_\Phi(\Phi) = \frac{\gamma}{\Gamma(\gamma)}
\left( \frac{\gamma\Phi}{\ave{\Phi}} \right)^{\gamma-1}
\exp\left( - \frac{\gamma\Phi}{\ave{\Phi}} \right) ,
\end{equation}
where the scale parameter is given by $\ave{\Phi}/\gamma$ and
the shape parameter is $\gamma=\ave{\Phi}^2/\Phirms^2$. It
should be noted that there are no fit parameter when comparing
this prediction to experimental measurements. The above equation
is exactly the form of the Gamma distribution found empirically to
describe plasma fluctuations in the \sol\ of TCV across a broad
range of plasma parameters \cite{graves}. Recently, it has also
been shown to describe fluctuations recorded by gas puff imaging
measurements in the \sol\ of the Alcator C-Mod tokamak
\cite{garcia-psi}.

In summary, a stochastic model for intermittent fluctuations in the
boundary region of magnetized plasmas has been constructed as a
random sequence of bursts which represent radial motion of blob-like
structures. The mean
plasma density is given by the average burst amplitude and the
ratio of burst duration and waiting times. In the case of exponentially
distributed burst amplitudes and waiting times, the amplitude probability
density function is shown to by a Gamma distribution. This simple model
thus explains the salient fluctuation statistics found in numerous
experimental measurements and elucidates the role of burst statistics
for large \sol\ plasma densities and fluctuation levels. The general
parabolic relation between skewness and kurtosis moments
predicted by this model likely explains the wide spread observation
of this scaling relation in physical systems dominated by intermittent
fluctuations.

Discussions with B.~La{B}ombard, M.~Melzani, H.~L.~P{\'e}cseli, R.~A.~Pitts, M.~Rypdal and J.~L.~Terry are gratefully acknowledged.

\clearpage

\end{document}